\documentclass[letterpaper, 10 pt, conference]{ieeeconf}  

\IEEEoverridecommandlockouts                              

\usepackage{xspace}
\newcommand{\umax}{\ensuremath{u_\mathrm{max}}\xspace}
\newcommand{\gamstate}{\ensuremath{\gamma_\mathrm{ctrl}}\xspace}
\newcommand{\gamlearn}{\ensuremath{\gamma_\mathrm{learn}}\xspace}
\newcommand{\tinp}{\ensuremath{t_\mathrm{inp}}\xspace}
\usepackage{cite}
\usepackage{inputenc}
\usepackage{balance}
\usepackage{xspace} 
\usepackage{url}
\usepackage[caption=false,font=footnotesize]{subfig}
\usepackage{paralist}
\usepackage{amsmath,amssymb}
\usepackage{mathtools}
\usepackage{xfrac}
\usepackage{array}
\usepackage{algorithm,algorithmic}
\usepackage{textcomp} 
\usepackage{siunitx}
\usepackage{pgfplots}
\usepackage{xcolor}
\pgfplotsset{compat=newest,unit code/.code={\si{#1}},plot coordinates/math parser=false,grid style={gray}}
\usepgfplotslibrary{units,external,groupplots}
\usetikzlibrary{positioning,angles,quotes,patterns,shapes,arrows,shadows,backgrounds}
\usepackage{csquotes}

\newtheorem{theo}{Theorem}

\newtheorem{remark}{Remark}
\newtheorem{example}{Example}

\DeclareMathOperator*{\E}{\mathbb{E}}

\newcommand{\norm}[1]{\left\lVert#1\right\rVert}
\newcommand{\abs}[1]{\left\lvert#1\right\rvert}
\newcommand*\diff{\mathop{}\!\mathrm{d}}
\DeclareMathOperator*{\R}{\mathbb{R}}

\let\originalleft\left
\let\originalright\right
\renewcommand{\left}{\mathopen{}\mathclose\bgroup\originalleft}
\renewcommand{\right}{\aftergroup\egroup\originalright}

\newcommand\figref[1]{Fig.~\ref{#1}}
\newcommand\tabref[1]{Table~\ref{#1}}
\newcommand\secref[1]{Sec.~\ref{#1}}

\newcommand{\eg}{\emph{e.g.},\xspace}
\newcommand{\ie}{\emph{i.e.},\xspace}

\usepackage{ifthen} 
\newboolean{authnotes} 

\setboolean{authnotes}{true} 

\ifthenelse{\boolean{authnotes}}
{
\newcommand{\db}[1]{\footnote{{\bf\color{blue} Dominik: #1}}}
\newcommand{\fs}[1]{\footnote{{\bf\color{orange} Friedrich: #1}}}
\newcommand{\kj}[1]{\footnote{{\bf\color{green!50!black} Kalle: #1}}} 
\newcommand{\st}[1]{\footnote{{\bf\color{green!50!black} Sebastian: #1}}}
}
{
\newcommand{\db}[1]{}
\newcommand{\fs}[1]{}
\newcommand{\kj}[1]{}
\newcommand{\st}[1]{}
}
\tikzstyle{block} = [draw, rectangle, minimum height=2em, minimum width=3em]
\tikzstyle{addon} = [draw, rectangle, rounded corners]
\tikzstyle{pinstyle} = [pin edge={<-,thin,black}]
\tikzstyle{pinstyle2} = [pin edge={->,thin,black}]
\tikzstyle{mult} = [draw, isosceles triangle]
\tikzstyle{circ} = [draw, circle]
\tikzstyle{coord} = [coordinate]
\tikzstyle{circ2} = [draw, circle,minimum width=3pt, inner sep=0]
\tikzset{>=latex}
\tikzset{radiation/.style={{decorate,decoration={expanding
waves,angle=90,segment length=4pt}}}}
\usepackage{hyphenat}

\title{\LARGE \bf
Event-triggered Pulse Control with Model Learning (if Necessary)
}

\author{Dominik Baumann$^{1,2}$, Friedrich Solowjow$^{1}$, Karl Henrik Johansson$^{2}$, and Sebastian Trimpe$^{1}$
\thanks{$^{1}$Intelligent Control Systems Group, Max Planck Institute for Intelligent Systems, Stuttgart/T\"{u}bingen, Germany. 
Email: dbaumann@tuebingen.mpg.de, fsolowjow@is.mpg.de, trimpe@is.mpg.de}%
\thanks{$^{2}$Division of Decision and Control Systems, KTH Royal Institute of Technology, Stockholm, Sweden.
Email: kallej@kth.se}%
\thanks{This work was supported in part by the German Research Foundation (DFG) within the priority program SPP 1914 (grant TR 1433/1-1),
         the Cyber Valley Initiative, the IMPRS-IS, and the Max Planck Society.}%
}

\usepackage{fancyhdr}
\newcommand{\mytitle}{\textbf{Accepted final version.}
To appear in \textit{Proc. of the American Control Conference (ACC), 2019}.\\
\copyright 2019 AACC.}
\begin{document}

\maketitle
\thispagestyle{fancy}	
\pagestyle{empty}


\begin{abstract}
    In networked control systems, communication is a shared and therefore scarce resource.
    Event-triggered control (ETC) can achieve high performance control with a significantly reduced amount of samples compared to classical, periodic control schemes.
    However, ETC methods usually rely on the availability of an accurate dynamics model, which is oftentimes not readily available.
    In this paper, we propose a novel event-triggered pulse control strategy that learns dynamics models if necessary. 
    In addition to adapting to changing dynamics, the method also represents a suitable replacement for the integral part typically used in periodic control.
\end{abstract}

\section{Introduction}
In modern engineering, control systems are often connected over communication networks.
Common examples for these networked control systems (NCS) include automation industry, where multiple plants have to be controlled often by a remote controller, building automation, where sensors and actuators are deployed to regulate the indoor climate, and coordinated flight of a swarm of drones. 
Communication then is a shared and therefore limited resource, thus, traditional, periodic control approaches are not feasible.

In this paper, we present a new architecture for event-triggered pulse control that quantifies model accuracy and, if necessary, automatically identifies system dynamics through learning.
The decision, whether learning is necessary, is taken by a learning trigger.
The proposed framework can cope with changing dynamics and load disturbances, that way replacing the integrator from periodic control.

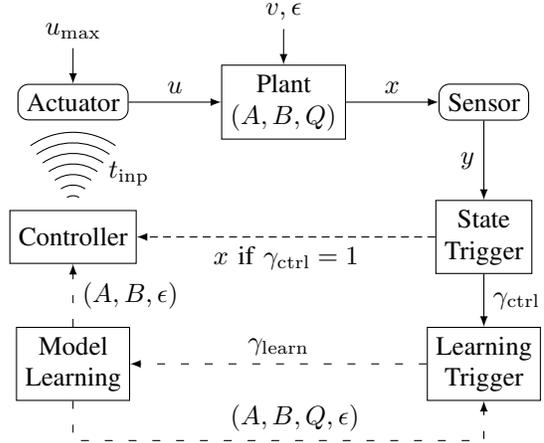
\begin{figure}
     \centering
     \begin{tikzpicture}
\node[block, pin={[pinstyle]above:$v,\epsilon$}, node distance=2em,align=center,fill=white](plant){Plant\\ $(A,B,Q)$};
\node[addon, right = 3.5em of plant](sens){Sensor};
\node[addon, left = 3.5em of plant, pin={[pinstyle]above:\umax}](act){Actuator};
\node[block, below = 3em of sens, align = center](et){State\\Trigger};
\node[block]at(et-|act)(ctrl){Controller};
\node[block, below = 2em of et, align = center](et2){Learning\\Trigger};
\node[block, align = center]at(et2-|ctrl)(ml){Model\\Learning};

\draw[->] (act) --node[midway,above]{$u$} (plant);
\draw[->] (plant) -- node[midway,above]{$x$} (sens);
\draw[->] (sens) --node[midway,left]{$y$} (et);
\draw[->, densely dashed] (et) -- node[midway,below]{$x$ if $\gamstate = 1$}(ctrl);
\draw[->] (et) --node[midway, right]{\gamstate} (et2);
\draw[->,loosely dashed] (et2) --node[midway,above]{\gamlearn} (ml);
\draw[->, loosely dashed] (ml) --node[midway,right]{$(A,B,\epsilon)$} (ctrl);
\draw[radiation,decoration={angle=35}] ([shift={(0,+0.05cm)}]ctrl.north)-- node[coordinate, pos=0.4](tinp){} ([shift={(0,-0.05cm)}]act.south);
\node[right=1em of tinp]{\tinp};
\draw[->, loosely dashed] (ml.south) -- +(0,-1.5em) -| node[pos=0.27,above]{$(A,B,Q,\epsilon)$} (et2);

\end{tikzpicture}
     \caption{Block diagram of the proposed control design. Dashed lines represent connections that are only active in case of an event.}
     \label{fig:blockdiag}
 \end{figure}

A block diagram of the approach is provided in \figref{fig:blockdiag}.
We consider a plant with sensors and actuators, subject to process noise and disturbances ($v$ and $\epsilon$), and input saturations \umax.
Controller and actuator are connected over a communication network.
Since communication is a scarce resource, periodic communication is not desirable and, therefore, we employ an event-triggered design (block \enquote*{State Trigger}).
In case of an event, we apply a pulse with length \tinp to reset the system to its equilibrium state.
The pulse length naturally depends on the system dynamics.
To obtain an accurate model of the system dynamics, we leverage system identification techniques to learn the model from data.
As learning may be expensive, \eg due to the involved computations, we only learn a new model if necessary, for instance, in case of a poor initial model or if the dynamics have changed.
This decision process is made by the \enquote*{Learning Trigger}.
Based on a statistical analysis of the time between events, the learning trigger decides whether the model of the system dynamics is accurate enough.
If not, learning of a new model is triggered.
We thus have two different triggers: the state trigger (\gamstate), which triggers communication of control commands if necessary, and a learning trigger (\gamlearn), which triggers learning in case of bad performance.

\emph{Contributions:} We make the following contributions:
\begin{itemize}
    \item A new architecture for event-triggered pulse control that automatically learns dynamics model to cope with changing dynamics;
    \item Development of a learning trigger for ETC, which allows to automatically identify system dynamics if necessary;
    \item Handling load disturbances by learning and compensating for them, thus replacing the integrator typically used in periodic control in a way suitable for ETC.
\end{itemize}

\emph{Related work:}
Reducing communication is one key aspect of NCS and has been addressed by introducing event-triggered methods~\cite{HeJoTa,Mi15}.
While most of these approaches are based on a dynamical model of the system, the model is typically assumed to be given and not learned from data as proposed herein.

The \enquote*{Learning Trigger} in \figref{fig:blockdiag} is based on a framework developed in recent work~\cite{so18}.
There, ETL was introduced to trigger learning experiments in event-triggered state estimation (ETSE).
Here, we extend this framework to control. 
We look at a straightforward type of ETC, namely event-triggered pulse control. 
In contrast to~\cite{so18}, which considered a linear system perturbed only by Gaussian noise, we also consider load disturbances herein.

Using Dirac inputs for ETC, as we do for developing the learning trigger, has also been investigated in other works on ETC such as~\cite{astrom2008event,henningsson2008sporadic,meng2012optimal}.
However, the approximation we propose to take into account input saturations has not been discussed therein.
Moreover, none of these references considers learning approaches to cope with changing system dynamics or disturbances.

The problem of finding a replacement for continuous or periodic integral control that is suitable for ETC (\eg to deal with load disturbances) has for instance been addressed in~\cite{cervin2007limit}, where a disturbance observer is used.
Instead of introducing a disturbance observer, we directly include the load disturbances in the learning framework.
As PID-controllers are the most common controllers used in industry, event-triggered PID-control has also been investigated starting from~\cite{aaarzen1999simple}.
A particular problem here is the replacement of the integral part of the PID-controller~\cite{sanchez2012event}.
Mostly, a network between sensor and controller is considered, thus the main problem for the integral part is the non-constant sampling time of the event-triggered mechanism.
In~\cite{durand2009further}, this is dealt with by explicitly taking into account the actual sampling time instead of assuming a nominal, constant sampling time.
A different approach is presented in~\cite{rabi2008event}, where the event detector is connected to the sensor.
Instead of looking at the absolute value of the integrator, the difference between the current value and the value at the last triggering instant is used to trigger communication, as a constant value of the integrator indicates a control error of zero.
Replacing the integrator through model learning has  not been proposed yet.

Event-triggered controllers can also be learned from data without learning a model. 
Such approaches are proposed, for example, in~\cite{VaFe,ZhNiHeXu,YaHeLi17,baumann2018deep}.
In contrast to those approaches, we use a specific control design and use learning to obtain accurate dynamic models.

\emph{Outline:} In the following section, we formulate the problem setting.
After that, we introduce the approach for event-triggered pulse control and discuss the concrete implementation of the learning trigger in \secref{sec:dual_prob}.
In \secref{sec:num_exp}, we will present a numerical study and conclude with a discussion in \secref{sec:conclusions}.

\section{Problem Formulation}
\label{sec:essentials}
We consider linear, time-invariant systems of the form 
\begin{align}
\label{eqn:sys_gen}
    \diff x(t) = Ax(t)\diff t + Bu(t) \diff t + \epsilon\diff t + Q\diff W(t),
\end{align}
with the state $x(t)\in\R^n$, the control input $u(t)\in\R^m$, the constant load disturbance $\epsilon\in\R^n$, and $W(t)\in\R^n$ a multidimensional Wiener process representing process noise.
We assume that we can measure the full state, thus, $y=x$ in \figref{fig:blockdiag}.

As depicted in \figref{fig:blockdiag}, control commands have to be transmitted over a communication network.
We thus employ an event-triggered design, with the block \enquote*{State Trigger} implemented by
\begin{align}
\label{eqn:sample_time}
    \gamstate = 1 \iff \norm{x(t)}_2 \ge \delta,
\end{align}
where $\delta$ is a user-defined threshold and is essentially the deviation from the equilibrium that we are willing to tolerate.
In case of an event, \ie $\gamstate=1$, we apply a pulse to reset the system to its equilibrium,
\begin{align}
    \label{eqn:inp_gen}
    u(t) = \begin{cases}
    0 & \text{if } t-t_k> \tinp \lor t_k = 0 \\
    \phi_{\umax}\left(\hat{A},\hat{B},\hat{\epsilon},x_{t_k}\right) & \text{if }t-t_k\le \tinp \land t_k\neq 0,
    \end{cases}
\end{align}
where $\phi_{\umax}$ is the pulse generating policy, $(\hat{A}, \hat{B}, \hat{\epsilon})$ captures the model of the system dynamics, $x_{t_k}$ is the state of the system at the triggering instant $t_k$, $\tinp$ is the pulse length (see \figref{fig:blockdiag}), and \umax the maximum input the actuator can apply.
By applying a pulse with appropriate length, we can reset the system to its equilibrium state.
This, however, requires that we have a model that accurately describes the true system dynamics.
We obtain this model, and adapt it in case the dynamics change, via model learning techniques.
Model learning may be expensive due to the involved computations or required exploration, therefore, we only want to learn in case the estimated dynamics $(\hat{A}, \hat{B}, \hat{\epsilon})$ deviate too much from the true dynamics $(A,B,\epsilon)$. 
Since the true dynamics are unknown, the decision needs to be based on some implicit feature, which will be the communication signal.
Developing such a learning scheme for ETC is the main objective of this paper.

\section{Event-triggered Pulse Control with Model Learning}
\label{sec:dual_prob}
In this section, we present the event-triggered control and learning framework.
We start with a derivation of the learning trigger assuming Dirac impulse control, then show how we learn the system dynamics, and finally detail practical pulse control (with bounded pulses) for first-order systems.

\subsection{Event-triggered Learning for Control}
\label{sec:etl}
The learning trigger is based on the framework presented in~\cite{so18} for ETSE.
Here, we extend this framework to ETC.
For the theoretical analysis, we assume a control strategy based on Dirac impulses, \ie here, we will ignore the assumption of having an input saturation at the actuator.
Using Dirac impulses, we can, similarly as in ETSE~\cite{so18}, reset the error to zero at communication times.
We then have a control law of the form
\begin{align}
     \label{eqn:ctrl_inp}
     u(t) = F \delta_{t_k}(t),
\end{align}
where $F$ is the control gain and $\delta_{t_k}$ the Dirac impulse.
In particular, the control input is zero apart from the triggering times $t_k$, where $t_k$ corresponds to $\gamstate=1$ in~\eqref{eqn:sample_time}.
To further analyze this scheme, we write~\eqref{eqn:sys_gen} in integrated form,
 \begin{align}
     \label{eqn:sys_gen_int}
     \begin{split}
     x\left(t_k\right) &= \int\limits_{t_{k-1}}^{t_k}e^{A\left(t_k-t\right)}Bu(t)\diff t \\
     &+ \underbrace{\int\limits_{t_{k-1}}^{t_k}e^{A\left(t_k-t\right)}\epsilon\diff t+\int\limits_{t_{k-1}}^{t_k}e^{A\left(t_k-t\right)}Q\diff W(t)}_{\coloneqq N\left(t_k\right)} \\
     &= e^{A\left(t_k-t_k\right)}BF + N\left(t_k\right)\\
     &= BF + N\left(t_k\right),
     \end{split}
 \end{align}
 where we assume that the process starts in $x(t_{k-1})=0$ and $N(t_k)$ is the measurement we get before applying the impulse.
 By setting~\eqref{eqn:sys_gen_int} to zero we can (assuming $B$ invertible) show that $F=-B^{-1}N(t_k)$ resets the system to zero.
 Implementing such a control law then also fulfills the prior assumption of $x(t_{k-1}) = 0$, as the system starts in zero after every triggering instant.
 In \secref{sec:pulse_ctrl}, we will drop the assumption of being able to apply Dirac impulses as inputs and instead apply pulses with the maximum input \umax for a given time.
 
 Considering a control law as proposed in~\eqref{eqn:ctrl_inp}, we thus have a random process that always starts in zero.
 This is only true, if the input matrix $B$ is known exactly.
 In that case, the sole cause of an error would be propagated noise and the load disturbance $\epsilon$.
Therefore, in case of no communication, we obtain
 \begin{align}
    \label{eqn:etl_ctrl_err_2}
    x(t) = \int\limits_0^t e^{A(t-s)} \epsilon\diff s + \int\limits_0^t e^{A(t-s)} Q\diff W(s).
\end{align}

We can now define a stopping time $\tau$ as the first moment the state crosses the threshold $\delta$, which resets the error to zero,
\begin{align}
\label{eqn:etl_stop_ctrl}
\tau \coloneqq \inf \left\{t:\norm{x(t)}_2\ge \delta\right\}.
\end{align}

The stopping times defined in~\eqref{eqn:etl_stop_ctrl} coincide with the time between communication, hence, \enquote*{stopping times} and \enquote*{inter-communication times} will be used synonymously hereafter.
We can now further define the expected value of these stopping times, $\E[\tau \vert x(0) = 0]$, which is the average inter-communication time of the system.
This expected value can be obtained via Monte Carlo simulations (for a more detailed discussion, see~\cite{so18}).

If we had a perfect model of the system dynamics, the average inter-communication times that we observe in the system should approach the expected value of the stopping time.
If both values deviate by too much, we have evidence that the model is inaccurate and can trigger learning of a new model.
Precisely, we define the learning trigger in \figref{fig:blockdiag} as
\begin{align}
\label{eqn:etl_learn_trig}
\gamlearn = 1 \iff \abs{\frac{1}{N}\sum_{i=1}^N \tau_i - \E[\tau]} \ge \kappa.
\end{align}
In this equation, $\gamlearn\!=\!1$ indicates that a new model shall be learned, $\E[\tau]$ is approximated using Monte Carlo simulations, \ie $\E[\tau]\!\approx\!\frac{1}{M}\sum_{i=1}^{M}\tau_i^{\mathrm{sim}}$, and $\tau_1, \tau_2, \ldots, \tau_N$ define the last $N$ empirically observed inter-communication times.
Due to the randomness of the process, it can still happen that we trigger learning despite the model being perfect.
Assuming that the stopping times are bounded by $\tau_\mathrm{max}$, the confidence level can be quantified using Hoeffding's inequality~\cite{probability1963hoeffding} and influenced through the design parameter~$\eta$.
\begin{theo}
\label{thm:learn_trig}
Let the parameters $\eta$, $N$, $M>N$, and $\tau_\mathrm{max}$ be given, $\tau_1,\ldots,\tau_N$ and $\tau_1^\mathrm{sim},\ldots,\tau_M^\mathrm{sim}$ independent and identically distributed, and assume a perfect model.
For
\begin{align}
    \kappa = \tau_\mathrm{max}\sqrt{-\frac{2}{N}\ln{\frac{\eta}{4}}}
\end{align}
we obtain
\begin{align}
    \label{eqn:approx_learn_trig}
    \mathbb{P}\left[\left| \frac{1}{N}\sum\limits_{i=1}^{N}\tau_i - \frac{1}{M}\sum\limits_{i=1}^{M}\tau_i^{\mathrm{sim}} \right|\ge\kappa\right]<\eta.
\end{align}
\end{theo}
\vspace{1ex}
\begin{proof}
We compare stopping times obtained via Monte Carlo simulations with stopping times observed from the real process.
In both cases, we have a random process that always starts in zero.
This is the same setting as investigated in~\cite{so18}, thus, the theorem can be proven as shown therein.
\end{proof}

Boundedness of the stopping times can easily be ensured in practice by applying a control input the latest when $\tau_\mathrm{max}$ is reached.
The confidence parameter $\eta$ then basically defines the tradeoff between accepting an inaccurate model or triggering learning despite the model being perfect.
Intuitively, $\eta$ defines the probability that the error $\kappa$ is observed, while empirical and expected stopping times are drawn from the same distribution (\ie we have a perfect model).
If this probability is below a predefined threshold, we trigger learning of a new model.

\subsection{Model Learning}
\label{eqnLlearn_dist}
For the derivation of the stopping times as well as for the final controller design, we need knowledge of the full system dynamics (matrices $A$ and $B$) and the load disturbance $\epsilon$. 
To calculate the stopping times, \eg via Monte Carlo simulations, we additionally need knowledge of the process noise variance $Q$.
To estimate the model, we rewrite system~\eqref{eqn:sys_gen} in discrete time,
\begin{align}
\label{eqn:sysid_dist}
\begin{split}
    x(k+1) &= A_\mathrm{d}x(k) + B_\mathrm{d}u(k) + \epsilon + v(k) \\
    &= \begin{pmatrix}
    A_\mathrm{d}&B_\mathrm{d}&\epsilon\end{pmatrix}
    \begin{pmatrix}
    x(k) \\ u(k) \\ 1\end{pmatrix}+v(k),
\end{split}
\end{align}
with $A_\mathrm{d}$, $B_\mathrm{d}$ the discrete-time system and input matrix, respectively, and $v(k)$ the discrete-time process noise.
That way, we can learn the system dynamics with standard least-squares techniques, as we will demonstrate in \secref{sec:num_exp}.

Having knowledge of the load disturbances, we can incorporate them in the control design in \secref{sec:pulse_ctrl}.
This represents a suitable solution to replace the integral part of standard, periodic controllers.

\begin{remark}
Another problem that may be considered with this approach is the knowledge of the zero-level of the system.
We are considering an equilibrium at $x(t)=0$, but the measurements are actually voltage signals from a sensor and what zero means for that system is not clear from the beginning.
We can model this as a sensor bias, \ie we would have the following system dynamics
\begin{subequations}
\begin{align}
    x(k+1) &= A_\mathrm{d}x(k) + B_\mathrm{d}u(k) + v(k)\\
    y(k) &= x(k) + \xi,
\end{align}
\end{subequations}
where $\xi$ is the sensor bias.
Rewriting this yields
 \begin{align}
 \label{eqn:sysid_bias}
     \begin{split}
         y(k+1) &= x(k+1) + \xi \\
         &= A_\mathrm{d}\left(y(k)-\xi\right) + B_\mathrm{d}u(k) +v(k)\\
         &= \begin{pmatrix}
         A_\mathrm{d}&B_\mathrm{d}&\left(I-A_\mathrm{d}\right)\xi
         \end{pmatrix}
         \begin{pmatrix}
         y(k)\\u(k)\\1
         \end{pmatrix}+v(k).
     \end{split}
 \end{align}
Thus, we can identify the system dynamics and the sensor bias via least-squares techniques.
Estimating both, a sensor bias and a load disturbance, is, however, not possible, as they are not distinguishable given the output data. 
\end{remark}

\subsection{Implementation of Event-triggered Pulse Control with Input Saturation}
\label{sec:pulse_ctrl}
If the state of the system exceeds the threshold $\delta$, we want to quickly reset it to zero.
Application of Dirac impulses is not compatible with our assumption of an input saturation at the actuator.

Instead of applying Dirac impulses, we propose to apply the maximum input and vary the \emph{duration} of the pulse.
This has two main benefits: 1) The input cannot exceed the saturation and, thus, will drive the system state to its desired value, as it will not be limited by the saturation; 2) The system will be driven to zero as fast as possible, that way coming as close to the idealized Dirac input as possible.
This represents a straightforward approach to lift the idea of ETL to the ETC setting.
For the derivation of the length of the pulse, we will restrict ourselves to first-order systems (\ie $n=1$ in~\eqref{eqn:sys_gen} with scalar variables $a$ and $b$) and later comment on extensions to higher-order systems.

To derive the length of an impulse, we look at the system equation in integrated form.
If no event is triggered, \ie if the system is close to its desired state, we have $u(t) = 0$. 
At triggering times $t_k$ we apply the maximum input,
\begin{align}
\label{eqn:sys_int_umax}
        x(t) &= e^{at}x\left(t_k\right) + \int\limits_{t_k}^t e^{a\left(t-s\right)}\left(b\umax + \epsilon\right)\diff s.
\end{align}
The input shall be applied for long enough such that the state becomes zero.
We thus set~\eqref{eqn:sys_int_umax} to zero and solve for $t$ (setting $t_k=0$),
\begin{align*}
0 &\stackrel{!}{=} e^{at}x\left(0\right) + \int\limits_{0}^t e^{a\left(t-s\right)}\left(b\umax+\epsilon\right)\diff s,
\end{align*}
which leads to
\begin{align}
\label{eqn:inp_dur}
\tinp &= \frac{1}{a} \ln\left(\frac{b\umax+\epsilon}{ax\left(0\right)+b\umax+\epsilon}\right).
\end{align}

In~\eqref{eqn:sys_int_umax} and~\eqref{eqn:inp_dur}, we assumed the noise to be zero during the application of the pulse.
For the Dirac impulse, this holds, as the time of the application tends to zero.
Here, we explicitly derive how long the input will be applied, hence, the system will during this time also be excited by noise and we will not be able to exactly drive it to zero.
We take this into account when computing the stopping times.
Instead of starting at zero, we consider a process that starts in $x(0) \sim \mathcal{N}(0,\Sigma_0)$, with variance $\Sigma_0$.

\begin{remark}
\label{rem:higher-order}
For a system~\eqref{eqn:sys_gen} with state dimension $n>1$, a single pulse will generally not be sufficient to drive the state to zero.
Instead, we would have to change between maximum and minimum input, which leads to a bang-bang controller~\cite{bellman1956bang}.
Then, $\tinp$ in \figref{fig:blockdiag} would be a vector containing the switching points.
Different algorithms have been proposed in literature to find these switching points~\cite{kaya1996computations,wen1987algorithm}.
This paper is a first approach towards combining the ideas of ETL with ETC and the extension to bang-bang type of controllers is beyond the scope of this work.
Generalizations to such or other control structures are, however, certainly important questions for future work.
\end{remark}

\section{Numerical Study}
\label{sec:num_exp}
For the numerical study, we will consider a collection of first-order processes.
For each system, we assume a remote controller that is colocated with the sensor, but needs to transmit its actuation commands over a communication network, where all controllers share the same network.
For all examples, we assume a threshold of $\delta=0.02$.
As parameters of the learning trigger, we choose a confidence level $\eta = 0.05$, $N=\num{2000}$, $M=\num{10000}$, and $\tau_\mathrm{max}=\SI{1}{\second}$.
According to Theorem~\ref{thm:learn_trig}, we then obtain $\kappa \approx \num{0.066}$.

In~\cite[p.~227]{aastrom2006advanced}, a batch of process models that are common in process industry is collected.
Among others, typical parameters for first-order systems with time delay are provided.
We will start the numerical investigation of the proposed framework in~\secref{sec:stable} by looking at these models, but we neglect the time delay.
The models have stable dynamics as is a common property in process industry.
To showcase the capability of the framework to also deal with unstable systems, we will consider such examples in \secref{sec:unstable}.
For all investigated systems, we will consider additional process noise and load disturbances.
We model the load disturbance to enter with the input, similar as for instance done in~\cite[p.~54]{aastrom1995pid}.

As in~\eqref{eqn:sys_gen}, we assume continuous-time systems
\begin{align}
    \label{eqn:nom_dyn}
    \begin{split}
    \diff x(t) &= ax(t)\diff t + b\left(u(t)+\epsilon\right)\diff t + Q\diff W(t),
    \end{split}
\end{align}
which we discretize with a sample time of \SI{1}{\milli\second}.
The sample time is not equal to the update interval of the communication system and is only limited by the maximum frequency of the timers in the processors used for controller and actuator.
A fine discretization is necessary, as we will derive a continuous pulse length.
The finer the discretization is, the more accurate is the application of the pulse (and the earlier we notice if the system is outside the tolerable range).

\subsection{Stable Dynamics}
\label{sec:stable}
As a proof of concept, we first consider one specific from~\cite{aastrom2006advanced} (note that the first-order systems provided therein always lead to $a=b$):
\begin{example}
\label{exp:stable_single}
System~\eqref{eqn:nom_dyn} with $a=b = -0.01$, process noise $Q=10^{-4}$, load disturbance $\epsilon=5$, and maximum input $\umax = 100$.
In case learning is triggered, we always collect data for \SI{200}{\second} and then use these data points to identify the system dynamics.
\end{example}

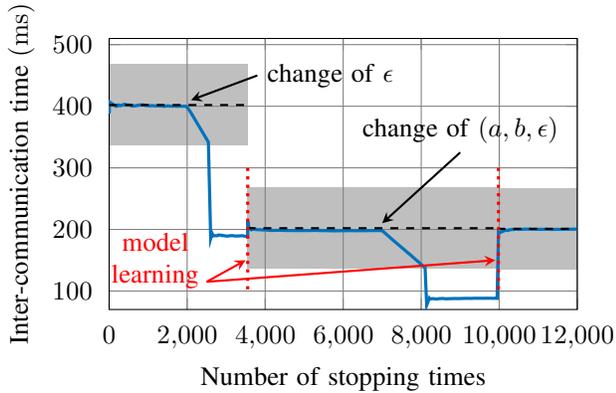
\begin{figure}
    \centering
%
%
\definecolor{mycolor1}{rgb}{0.00000,0.44700,0.74100}%
\begin{tikzpicture}
\pgfdeclarelayer{background}
\pgfsetlayers{background,main}

\begin{axis}[%
width=0.35\textwidth,
height=0.15\textheight,
scale only axis,
xmajorgrids,
ymajorgrids,
scaled ticks=false, 
tick label style={/pgf/number format/fixed,
 /pgf/number format/precision=5},
  ticklabel shift={.1cm},
at={(0.78in,2.684in)},
xlabel = Number of stopping times,
ylabel  = Inter-communication time,
unit markings = parenthesis,
y unit = ms,
xmin = 0,
xmax=12000,
ymin = 70,
ymax = 510
]
\begin{pgfonlayer}{background}
\fill[lightgray](0,468)rectangle(3548,336);
\fill[lightgray](3549,136) rectangle(9953,268);
\fill[lightgray](9954,135) rectangle(12000,267);
\end{pgfonlayer}
\addplot [color=mycolor1, forget plot,very thick]
  table[row sep=crcr]{%
1	405\\
2	389\\
4	407\\
113	402.681415929204\\
173	399.924855491328\\
200	400.120000000003\\
241	402.116182572616\\
284	400.897887323947\\
365	402.646575342464\\
410	403.248780487804\\
457	401.789934354485\\
496	402.108870967742\\
565	401.145132743361\\
573	401.369982547993\\
590	400.46101694915\\
671	400.675111773475\\
755	401.829139072848\\
809	402.316440049442\\
841	401.683709869205\\
868	402.261520737324\\
934	401.29443254818\\
944	401.544491525427\\
1048	400.697519083973\\
1086	401.034990791894\\
1199	400.49791492911\\
1248	400.888621794875\\
1320	400.425757575758\\
1344	400.450892857145\\
1401	399.842255531767\\
1414	399.89674681754\\
1489	400.489590329082\\
1522	400.55387647832\\
1583	400.097283638657\\
1637	400.039095907145\\
1679	400.315664085763\\
1696	400.24174528302\\
1734	399.895040369091\\
1816	400.308920704847\\
1863	399.913043478264\\
1896	400.097046413503\\
1979	399.830217281458\\
2073	391.986499999999\\
2167	382.016000000003\\
2261	372.015500000001\\
2356	361.868499999997\\
2450	351.809999999998\\
2544	341.892999999996\\
2603	185.5\\
2639	193.57894736842\\
2666	189.784615384619\\
2776	189.811428571425\\
2800	190.974874371859\\
2868	190.782771535582\\
2902	189.887043189367\\
2921	190.128125000003\\
3022	189.083135391927\\
3108	189.814595660748\\
3165	189.914893617024\\
3231	189.314285714288\\
3310	189.393511988717\\
3422	189.817295980509\\
3550	188\\
3552	213.333333333336\\
3586	202.189189189186\\
3647	199.336734693876\\
3683	200.746268656716\\
3765	199.787037037036\\
3773	200.053571428572\\
3828	199.494623655912\\
3864	199.609523809522\\
3931	199.157068062828\\
3997	199.354910714283\\
4053	198.882936507936\\
4166	198.304700162073\\
4222	198.546805349186\\
4241	198.407514450868\\
4379	198.687951807231\\
4420	198.366245694604\\
4479	198.45483870968\\
4505	198.274058577408\\
4553	198.318725099605\\
4614	198.183098591551\\
4648	198.275705186534\\
4809	198.071428571428\\
4896	198.238307349668\\
4923	198.086608442507\\
4993	198.271468144041\\
5038	198.087306917398\\
5284	198.408645533142\\
5451	198.094637223978\\
5655	197.964\\
5684	197.805999999997\\
5863	197.803999999996\\
5989	198\\
6031	197.894\\
6155	198.129999999997\\
6314	197.879999999997\\
6422	198.057000000001\\
6479	197.919500000004\\
6600	198.018499999998\\
6640	198.146999999997\\
6784	198.152000000002\\
6922	198.218999999997\\
6954	198.054499999998\\
6997	198.127999999997\\
7064	194.582000000002\\
7158	189.281000000003\\
7252	184.031999999999\\
7346	178.940000000002\\
7440	173.915500000003\\
7534	168.751499999998\\
7629	163.688000000002\\
7723	158.697\\
7817	153.445\\
7911	148.246500000001\\
8005	143.022499999999\\
8099	137.879999999997\\
8140	77\\
8207	88.0588235294126\\
8215	87.2236842105267\\
8305	87.668674698798\\
8391	87.0079365079364\\
8471	87.7018072289138\\
8490	87.5840455840444\\
8542	87.9950372208405\\
8627	87.3872950819641\\
8864	87.8620689655145\\
8948	87.754017305313\\
9143	88.1354581673295\\
9293	87.9896013864854\\
9360	88.2538902538872\\
9625	88.2133243607022\\
9661	88.3055190538798\\
9946	88.1411178749331\\
9980	195.307692307695\\
10000	194.17391304348\\
10070	196.922413793101\\
10077	196.585365853658\\
10170	198.685185185182\\
10219	198.298113207544\\
10253	199.35117056856\\
10265	199.241157556273\\
10340	200.150259067355\\
10360	199.768472906406\\
10421	200.199143468948\\
10462	200.305118110235\\
10487	200.045028142587\\
10602	200.2237654321\\
10631	200.553914327917\\
10652	200.398280802292\\
10701	200.606425702812\\
10741	200.554002541299\\
10879	200.179459459461\\
10924	200.324742268043\\
11038	200.153136531364\\
11113	200.272648835205\\
11140	200.13069139966\\
11210	200.319267515923\\
11343	199.987760979122\\
11526	200.166666666664\\
11548	200.040777917187\\
11966	200.5\\
12063	200.724999999999\\
12233	200.513500000001\\
12262	200.582000000002\\
12378	200.413\\
12625	200.391499999998\\
12762	200.5965\\
12784	200.498\\
13144	200.805\\
13302	200.735500000003\\
13372	200.934999999998\\
13413	200.853499999997\\
13471	201.008999999998\\
13515	200.816500000001\\
13564	200.982499999998\\
13597	200.7935\\
13662	200.889000000003\\
13787	200.693500000001\\
13846	200.803999999996\\
13972	200.550999999999\\
14219	200.716500000002\\
14262	200.601999999999\\
14320	200.656999999999\\
14408	200.820500000002\\
14507	200.618999999999\\
14602	200.851999999999\\
14708	200.575499999999\\
14726	200.680500000002\\
14879	200.620999999999\\
14922	200.803500000002\\
15536	200.296999999999\\
15569	200.219499999999\\
15619	200.402999999998\\
15653	200.252\\
15770	200.339\\
15823	200.424500000001\\
15857	200.279999999999\\
15957	200.408499999998\\
16012	200.264999999999\\
};
\addplot[dashed,black,thick]coordinates{(0,402) (3548,402)};
\addplot[dashed,black,thick]coordinates{(3549,202) (9953,202)};
\addplot[dashed,black,thick]coordinates{(9954,201) (12000,201)};
\draw[stealth-,thick,black](axis cs: 2100,410)--(axis cs:3800,450)node[right]{change of $\epsilon$};
\draw[stealth-,thick,black](axis cs:7000,210) -- (axis cs: 9000, 325) node[above]{change of $(a,b,\epsilon)$};
\addplot[dotted,red,very thick]coordinates{(3552,300) (3552,100)};
\addplot[dotted,red,very thick]coordinates{(9980,300) (9980,100)};
\draw[stealth-,thick,red](axis cs:3500,150) -- (axis cs:2500,115);
\draw[stealth-,thick,red](axis cs:9900,150) -- (axis cs:2500,115);
\node at (axis cs:1175,150) [align = center,red] {model\\learning};
\end{axis}
\end{tikzpicture}%
    \caption{Average inter-communication times during one simulation for Example~\ref{exp:stable_single}. The solid line shows the empirically observed inter-communication times computed as a moving average over \num{2000} stopping times. The moving average is reset in case learning is triggered and when a new model has been learned. The dashed line indicates the expected inter-communication times with confidence interval $\pm \kappa$ highlighted in gray. The dynamics change after \num{2000} and after \num{7000} stopping times. 
    The points in time where new system matrices are learned are marked with vertical, red, dotted lines.
    In both cases, this causes a decrease in the inter-communication time (\ie more communication). 
    The inter-communication time increases again after learning new matrices.}
    \label{fig:avg_stop_times}
\end{figure}

In \figref{fig:avg_stop_times}, the average and expected inter-communication times are shown.
The average inter-communication times are computed with a moving average over \num{2000} stopping times, which we reset in case learning is triggered and after deriving new system matrices.
In the beginning, we assume that we have an accurate model, hence, the observed inter-communication times approach the expected ones.
After \num{2000} stopping times, we set the load disturbance to $\epsilon\!=\!10$.
As expected, the inter-communication times decrease and learning is triggered.
After learning (first vertical line in \figref{fig:avg_stop_times}), the empirical inter-communication times again approach the expected ones and we reduce communication.
In a second change, after \num{7000} stopping times, we have $\epsilon\!=\!20$ and $a=b=-0.05$.
Similar as before, this leads to a decrease of the average inter-communication time and learning is triggered.
Having learned new dynamics (second vertical line in \figref{fig:avg_stop_times}), the empirical inter-communication times again approach the expected ones, \ie average communication is reduced through learning.

We now consider the case where the initial matrices that we have of the process are wrong.
Both, nominal and true dynamics of the systems are again taken from~\cite{aastrom2006advanced}:
\begin{example}
\label{exp:stable_benchmark}
Two systems of the form~\eqref{eqn:nom_dyn} with nominal dynamics $a=b\in\{-0.1,-1\}$, load disturbances $\epsilon\in\{0.1,1\}$, $Q=10^{-4}$, and $\umax\in\{1,100\}$.
For the true dynamics, $a$ and $b$ are given in~\tabref{tab:stab_sys}, the noise variance and load disturbance are sampled from uniform distributions over the intervals $Q\in[10^{-4},10^{-3}]$, $\epsilon\in [0.1,0.2]$ (for system~1), and $\epsilon\in [1,5]$ (for system~2).
In case learning is triggered, we use all data collected so far to identify the system dynamics.
\end{example}

The inter-communication times before and after learning the dynamics are given in \tabref{tab:stab_sys}.
It can clearly be seen that learning helps in lowering communication.

\begin{table}[tbp]
    \centering
        \caption{Comparison of the average inter-communication times for Example~\ref{exp:stable_benchmark} before and after learning. System 1 is shown in the left column, system 2 in the right.}
    \begin{tabular}{cccccc}
        $a,b$ & Before & After  & $a,b$ & Before & After\\
        $-10$ & \SI{1}{\milli\second} & \SI{21}{\milli\second} & $-0.25$ & \SI{1}{\milli\second} & \SI{80}{\milli\second}\\
        $-\sfrac{4}{3}$ & \SI{1}{\milli\second} & \SI{39}{\milli\second} & $-0.05$ & \SI{69}{\milli\second} & \SI{132}{\milli\second}\\
        $-0.5$ & \SI{114}{\milli\second} & \SI{215}{\milli\second} & $-0.02$ & \SI{85}{\milli\second} & \SI{431}{\milli\second}\\
        $-0.25$ & \SI{111}{\milli\second} & \SI{428}{\milli\second} & $-0.01$ & \SI{62}{\milli\second} & \SI{384}{\milli\second}\\
        $-\sfrac{1}{6}$ & \SI{101}{\milli\second} & \SI{321}{\milli\second} & $-0.005$ & \SI{115}{\milli\second} & \SI{850}{\milli\second}
    \end{tabular}
    \label{tab:stab_sys}
\end{table}

\subsection{Unstable Dynamics}
\label{sec:unstable}

For the investigation of systems with unstable dynamics, we look at the following system:
\begin{example}
\label{exp:unst_benchmark}
System~\eqref{eqn:nom_dyn} with nominal dynamics $a=5$, $b=3$, $\epsilon=0.01$, $Q=10^{-4}$, and maximum input $\umax=1$.
The true dynamics are sampled from random distributions over the intervals $a \in [1,10]$, $b\in [1,2]$, $\epsilon \in [0.01, 0.02]$, and $Q\in[10^{-4},10^{-3}]$.
As in Example~\ref{exp:stable_benchmark}, we use all data we have collected so far to identify a new model in case learning is triggered.
\end{example}

In \tabref{tab:unst_sys}, we compare the average inter-communication times of all systems before and after deriving new system matrices.
For all of them, we observe a significant increase in the inter-communication times after learning, \ie communication is reduced. 
This demonstrates that the approach is also suitable for unstable systems.

\begin{table}[tbp]
    \centering
        \caption{Comparison of the average inter-communication times for Example~\ref{exp:unst_benchmark} before and after learning.}
    \begin{tabular}{cccccc}
        System & Before & After  & System & Before & After\\
        1 & \SI{44}{\milli\second} & \SI{239}{\milli\second} & 6 & \SI{128}{\milli\second} & \SI{271}{\milli\second}\\
        2 & \SI{136}{\milli\second} & \SI{349}{\milli\second} & 7 & \SI{56}{\milli\second} & \SI{240}{\milli\second}\\
        3 & \SI{107}{\milli\second} & \SI{283}{\milli\second} & 8 & \SI{55}{\milli\second} & \SI{219}{\milli\second}\\
        4 & \SI{135}{\milli\second} & \SI{256}{\milli\second} & 9 & \SI{89}{\milli\second} & \SI{279}{\milli\second}\\
        5 & \SI{163}{\milli\second} & \SI{363}{\milli\second} & 10 & \SI{34}{\milli\second} & \SI{273}{\milli\second}
    \end{tabular}
    \label{tab:unst_sys}
\end{table}

\begin{figure}
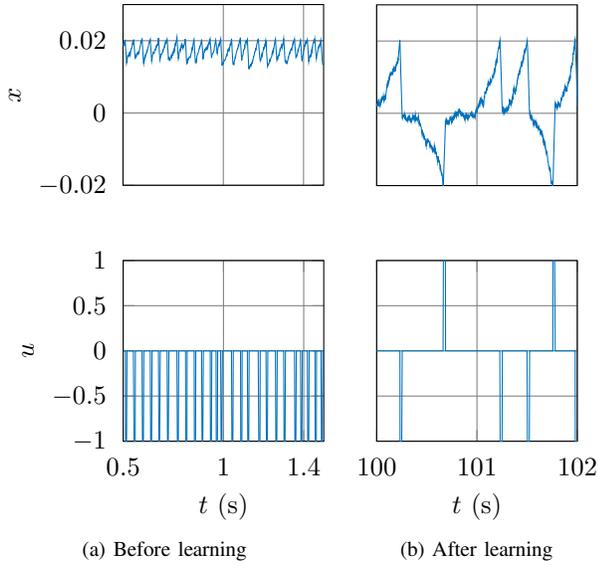

    \centering
    \subfloat[Before learning]{
    \input{Tikz/unst_sys_500-1500.tex}
    \label{sfig:uns_sys_before}
    }
    \subfloat[After learning]{
    \input{Tikz/unst_sys_100000-101000.tex}
     \label{sfig:uns_sys_after}
    }
    \caption{Performance of one specific system of Example~\ref{exp:unst_benchmark} before (left) and after (right) learning. It can be seen that before learning, the pulses are too short and the system is not reset to zero, while after learning the pulse length is appropriate. Further, communication is significantly reduced through learning.}
    \label{fig:uns_sys}
\end{figure}

In \figref{fig:uns_sys}, one specific system from Example~\ref{exp:unst_benchmark} is shown before (\figref{sfig:uns_sys_before}) and after (\figref{sfig:uns_sys_after}) learning new system matrices.
Due to the error in the initial matrices, the system is not reset to zero with the pulses before learning and, thus, new control inputs have to be generated very frequently.
After learning, the pulse length is such that the system is successfully reset, which also results in increased inter-communication times, and, therefore, less communication.
This is especially emphasized as in \figref{sfig:uns_sys_before}, before learning, we show only \SI{1}{\second}, while in \figref{sfig:uns_sys_after} we show \SI{2}{\second} and still observe far less pulses.

The study reveals that the proposed architecture enables us to increase inter-communication times through learning.
We are able to learn system dynamics and subsequently reset the state of the system to zero in case it leaves its tolerable range.
Through learning load disturbances, the architecture is a suitable replacement for integral control in event-triggered settings.

\section{Conclusion}
\label{sec:conclusions}
In NCS, communication is a scarce and limited resource.
In this work, we presented a framework for event-triggered  pulse control for NCS.
Most common event-triggered control approaches rely on the availability of an accurate dynamics model.
Contrary to that, the proposed framework does not rely on this assumption, but uses model learning instead.
As learning is expensive (\eg due to the involved computations), we only learn if necessary using the ETL framework.
By observing the communication behavior, we quantify the accuracy of the model and trigger learning of a new model only in case the accuracy is not sufficient.
The presented control design respects input saturations and can also handle load disturbances, thus essentially replacing the integral part of common periodic controllers.

A numerical study demonstrates the applicability of the approach and the benefit of learning the system dynamics.
After learning, we observe a significant increase in the inter-communication time.
However, the presented examples are first-order systems.
While we have outlined how higher-order systems could be treated (Remark~\ref{rem:higher-order}), the actual extension to such systems is an interesting topic for future work.
Moreover, we assumed that we are able to perfectly measure the full state of the system.
Incorporating Gaussian measurement noise is already possible with the presented approach.
How to extend the ETL framework to partial state measurements is subject to ongoing research.

In this work, we proposed for the first time to trigger model learning experiments in ETC to adapt to changes in the dynamics.
The learning trigger compares the expected and the observed time between communication.
While this is an intuitive approach, in some cases this trigger does not detect disturbed models.
A more robust behavior can be achieved by triggering on the full distribution, \eg via a Kolmogorov-Smirnoff test, which is subject of current research~\cite{solowjow2019event}.

\bibliographystyle{IEEEtran}
\bibliography{IEEEabrv,ref}

\end{document}